\documentclass[sigconf,screen,nonacm]{acmart}

\setcopyright{none}

\setcopyright{acmlicensed}
\acmISBN{978-1-4503-XXXX-X/18/06}
\acmDOI{XXXXXXX.XXXXXXX}
\newcommand{\showURL}[1]{\unskip}

\usepackage{booktabs}
\usepackage{multirow}
\usepackage{amsmath}
\usepackage{mathtools}
\usepackage{enumitem}
\usepackage{tikz}
\usetikzlibrary{arrows.meta,positioning,calc,fit,backgrounds}
\usepackage{pgfplots}
\usepgfplotslibrary{groupplots}
\pgfplotsset{compat=1.18}

\definecolor{pitblue}{HTML}{2F6B9A}
\definecolor{pitfill}{HTML}{EAF3FA}
\definecolor{leakred}{HTML}{C95B5B}
\definecolor{leakfill}{HTML}{F9E9E6}
\definecolor{successgreen}{HTML}{2E7D5B}
\definecolor{successfill}{HTML}{E8F4EC}
\definecolor{quietblue}{HTML}{E8F1F8}
\definecolor{quietgray}{HTML}{F5F7F9}

\title{A Historical Corpus Is Not a Historical System: Auditing Hindsight Leakage in Stateful Data Discovery}

\author{Yixi Zhou}
\authornote{These authors contributed equally to this work.}
\affiliation{%
  \institution{Hong Kong Baptist University}
  \city{Hong Kong SAR}
  \country{China}
}
\email{yxzhou@comp.hkbu.edu.hk}

\author{Fan Zhang}
\authornotemark[1]
\affiliation{%
  \institution{The University of Tokyo}
  \city{Tokyo}
  \country{Japan}
}
\email{zhang-fan@g.ecc.u-tokyo.ac.jp}

\author{Sikun Wang}
\affiliation{%
  \institution{Tokyo University of Science}
  \city{Tokyo}
  \country{Japan}
}
\email{jc19546883@gmail.com}

\author{Yingfan Xu}
\affiliation{%
  \institution{ShanghaiTech University}
  \city{Shanghai}
  \country{China}
}
\email{xuyf2024@shanghaitech.edu.cn}

\author{Haipeng Zhang}
\authornote{Corresponding author.}
\affiliation{%
  \institution{ShanghaiTech University}
  \city{Shanghai}
  \country{China}
}
\email{zhanghp@shanghaitech.edu.cn}

\acmConference[WSDM '27]{20th ACM International Conference on Web Search and Data Mining}{February 15--19, 2027}{Hong Kong, China}
\acmYear{2027}
\copyrightyear{2027}

\ccsdesc[500]{Information systems~Evaluation of retrieval results}
\ccsdesc[300]{Information systems~Information retrieval}

\begin{document}

\begin{abstract}
Offline replay should estimate what a discovery system could retrieve at a historical point, yet freezing the corpus leaves interaction memory unconstrained.  We formalize point-in-time (PIT) discovery through historical state $(D_t,\theta_t,M_{<i})$ and introduce a paired replay that changes only memory availability.  The protocol constructs PIT and full-stream Future views from behavior-only traces and audits selected entries with a Temporal Violation Rate.  Across three table--text domains, two stream regimes, two retrievers, and five seeds (216,000 rows), Future inflated Asset Recall@100 by 2.62--5.24 points; all 12 paired intervals excluded zero.  With behavior-only trace memory, PIT underperformed the no-memory Stateless condition; Future masked 32.7--48.4\% of that harm.  For a simulated positive-feedback cache, PIT added 4.65--18.96 points over Stateless while Future added another 4.11--9.58 points.  On five timestamped FreshStack topics, Future exceeded PIT by 2.72 points [1.75, 3.71].  Historical evaluation must version and validate memory with the corpus.
\end{abstract}

\keywords{point-in-time evaluation, temporal leakage, data discovery, stateful retrieval, offline replay}

\maketitle

\section{Introduction}
\label{sec:intro}

Researchers and practitioners use historical offline evaluation to estimate what a retrieval system could have returned at a past point in time.  Freezing the candidate corpus and retrieval index appears to reconstruct that system, but stateful discovery introduces a second time-dependent state: interaction memory.  Query traces, caches, profiles, and retrieval graphs may include observations that arrive after the replay anchor.  When an evaluator builds these states from the full log, post-anchor information changes the ranking even though every candidate asset passes the corpus-time filter.  The replay then measures a future-informed system over a historical corpus.

\begin{figure}[t!]
  \centering
  \includegraphics[width=\linewidth]{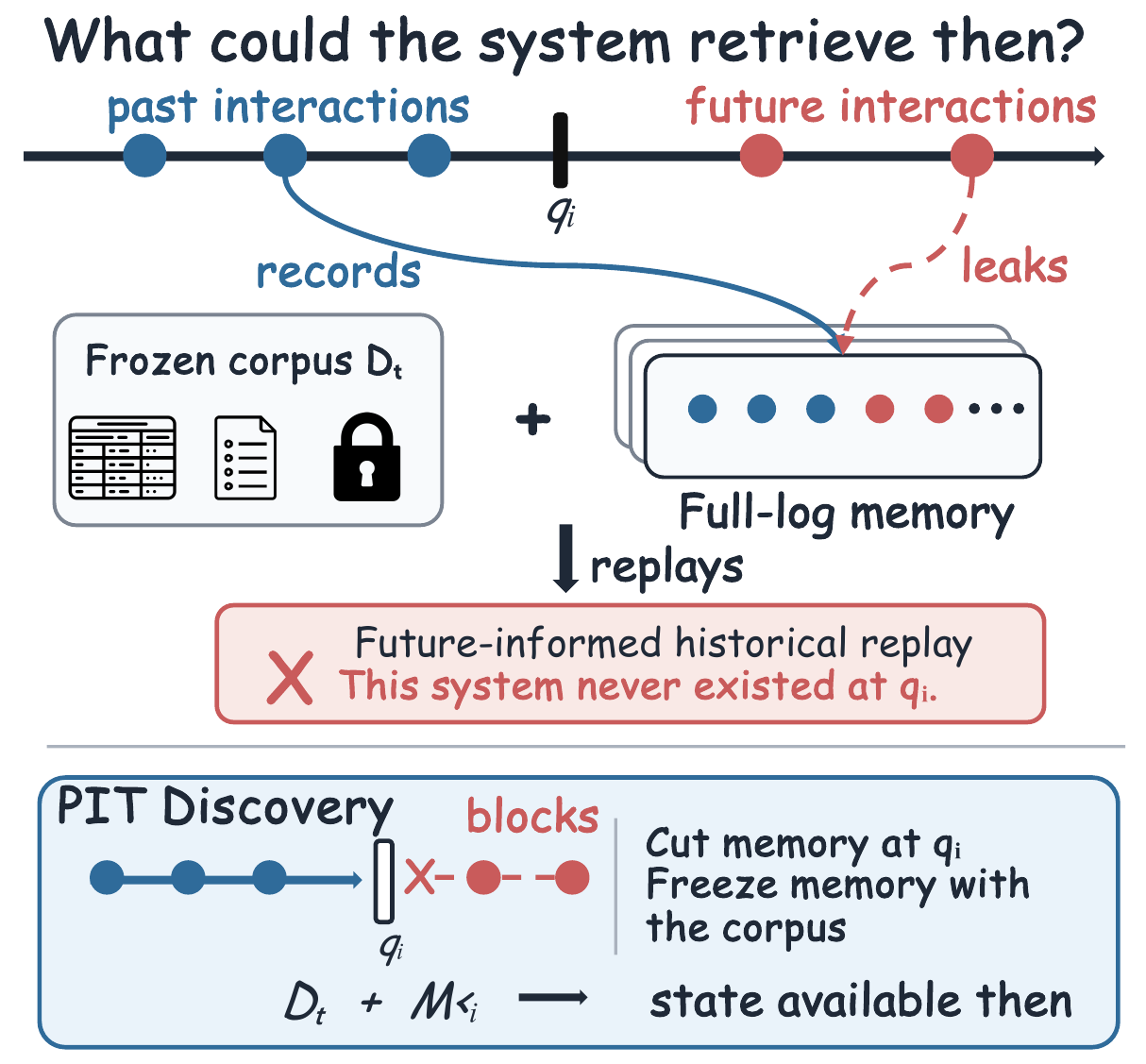}
  \caption{Freezing the corpus does not freeze historical system state.  Full-log memory leaks post-anchor interactions into a replay, while PIT Discovery binds the frozen corpus $D_t$ to memory $M_{<i}$ available before query $q_i$.}
  \Description{A timeline separates past and future interactions at anchor query q sub i. Full-log memory leaks future interactions into a frozen-corpus replay, while PIT Discovery blocks post-anchor traces and combines the frozen corpus with memory available before the anchor.}
  \label{fig:problem}
\end{figure}

Stateful discovery makes this gap consequential.  Multi-asset benchmarks measure whether systems retrieve coherent products spanning tables and text~\cite{zhang2025dpbench,feng2024cmdbench,choubey2025deepsearch}.  These protocols fix the corpus and ranking inputs, while deployed systems can reuse prior interactions.  Aggregate recall collapses temporal validity and retrieval quality into one score.  It therefore cannot reveal whether a gain came from historical evidence or later experience.

\begin{figure*}[t]
  \centering
  \includegraphics[width=\linewidth]{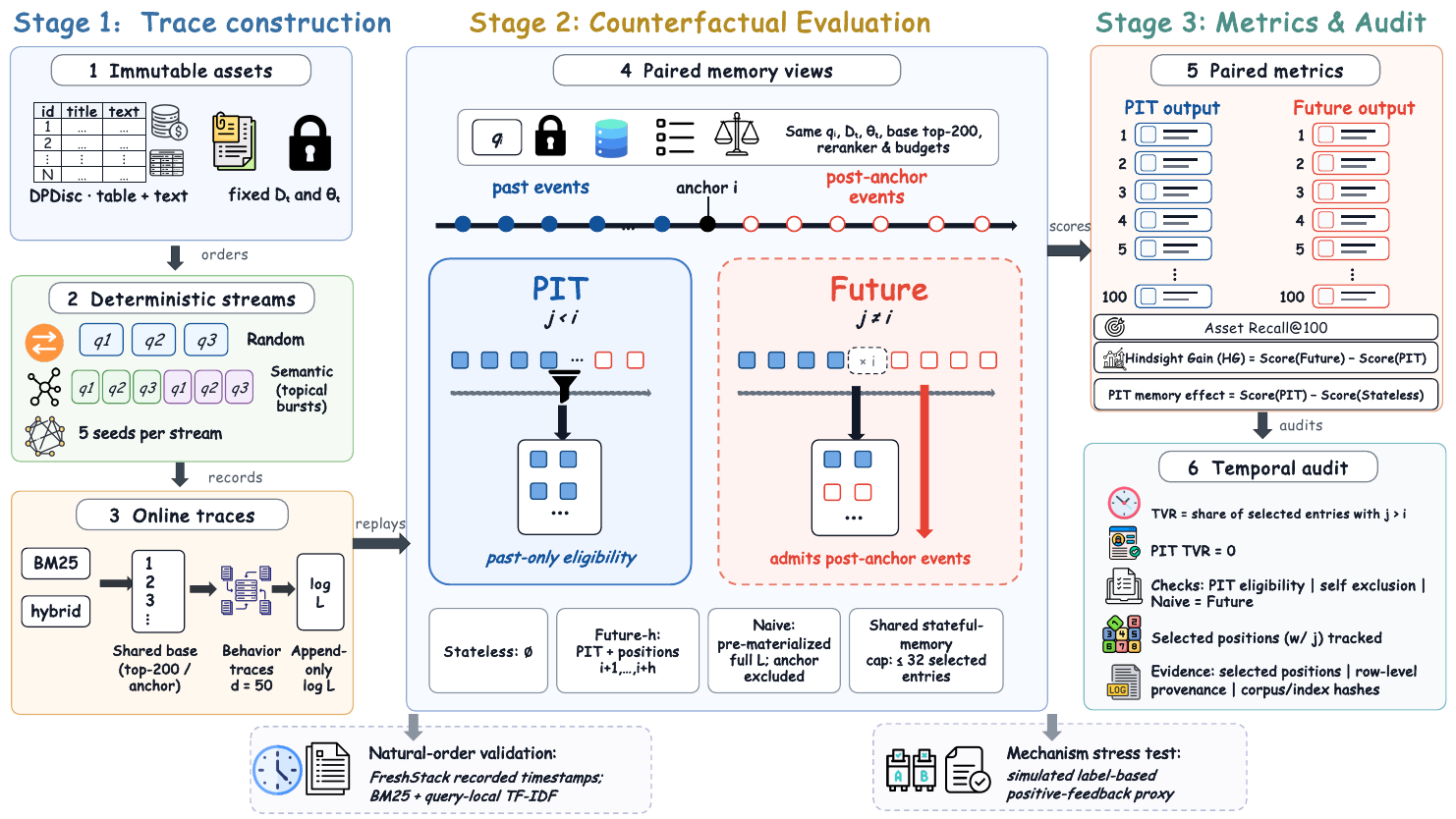}
  \caption{PIT Discovery constructs deterministic query streams and online behavior traces, compares matched PIT and Future memory views under fixed retrieval conditions, and reports paired quality metrics plus a temporal audit.}
  \Description{A three-stage pipeline builds immutable assets, deterministic streams, and online behavior traces. It compares matched PIT, Future, Stateless, Future-h, and Naive memory views under fixed retrieval conditions and a shared capacity limit. The pipeline computes Asset Recall at 100, Hindsight Gain, and PIT memory effect, then audits Temporal Violation Rate, eligibility, self-exclusion, selected memory positions, and provenance. Natural-order validation and a positive-feedback stress test appear below the main pipeline.}
  \label{fig:main}
\end{figure*}

We make the missing state variable explicit.  At corpus time $t$ and query-stream position $i$, a valid point-in-time (PIT) evaluation must bind the corpus snapshot $D_t$ to memory $M_{<i}$ induced only by observations available before the anchor.  This yields a clean counterfactual: keep the query, corpus, index, base ranking, reranking rule, and budgets identical, then vary only the temporal eligibility of memory.  We call the causal condition \emph{PIT} and the full-stream counterfactual \emph{Future}.  Their score difference is \emph{Hindsight Gain} (HG), which attributes the gap directly to future experience instead of dated corpora or retrained models.

Measuring this gap presents three challenges.  First, the evaluator must reconstruct interaction state without using relevance labels or the anchor's own trace.  Second, the evaluator must give PIT and Future memory views equal capacity; otherwise a score change could be a budget effect.  Third, a quality metric alone cannot certify historical validity.  \emph{PIT Discovery} converts a fixed multi-asset benchmark into deterministic query streams and records behavior-only traces online.  It materializes PIT, Future-$h$ (PIT plus the next $h$ events), and Future views at matched anchors, then emits retrieval quality and a Temporal Violation Rate (TVR).  Immutable manifests and row-level memory positions make the causal boundary auditable.

The empirical result is stronger than a generic warning about leakage.  We test ConvFinQA, HybridQA, and TATQA across semantic and random streams, BM25 and hybrid retrieval, and five seeds.  Future raises Asset Recall@100 over PIT by \textbf{2.62--5.24 percentage points}; all 12 query-cluster bootstrap intervals exclude zero.  The tested trace memory is harmful, and hindsight hides \textbf{32.7--48.4\%} of its deployable loss.  A simulated positive-feedback cache rules out an unhelpful memory rule as the sole cause: it produces $S<P<F$ in all 12 settings, with 4.11--9.58 points still attributable to hindsight.  A natural timestamp order over five FreshStack topics also yields a 2.72-point HG [1.75, 3.71].  Leakage can therefore distort the judgment about both harmful and useful memory under constructed and observed arrival orders.

This paper makes three contributions:
\begin{itemize}[leftmargin=*,nosep]
  \item \textbf{State formulation.} We define historical system state using $D_t$, $\theta_t$, and $M_{<i}$, and separate a legal historical memory snapshot from full-stream hindsight.
  \item \textbf{Paired protocol and audit.} We provide label-free stream construction, behavior-only traces, equal-budget memory views, future-horizon dosing, cryptographic manifests, and a zero-TVR validity invariant.
  \item \textbf{Measurement evidence.} We quantify systematic hindsight inflation, its locality and anchor-position boundaries, and its stability across retrievers, natural order, and two memory mechanisms, including a useful one.
\end{itemize}
Our evaluation claim permits different magnitudes for production logs and learned memories.  The conclusion holds across a harmful trace mechanism and a useful feedback proxy.  A score cannot represent historical capability unless every state that influenced it was historically available.

\section{Related Work}
\label{sec:related}

The representative protocols reviewed below vary several time-dependent components; none isolates interaction-memory availability under a fixed corpus.  Table~\ref{tab:related_axes} separates the four research lines that define our position.

\textbf{Data and evidence discovery.}
Recent benchmarks move retrieval beyond single-document fact lookup.  CMDBench studies coarse-to-fine retrieval across multimodal sources~\cite{feng2024cmdbench}; TARGET isolates table retrieval for generative tasks~\cite{ji2025target}; and Deep Search requires source-aware, multi-hop retrieval over heterogeneous artifacts~\cite{choubey2025deepsearch}.  DPBench/DPDisc maps each request to a coherent product containing multiple tables and passages~\cite{zhang2025dpbench}.  HybridQA, TAT-QA, and ConvFinQA provide the table--text evidence structures behind its three domains~\cite{chen2020hybridqa,zhu2021tatqa,chen2022convfinqa}.  These benchmarks expose asset retrieval and product completeness, while their standard protocols give each request a stateless retriever.  We retain DPDisc's product semantics and add an interaction-state axis.

\textbf{Stateful and conversational retrieval.}
Conversational search establishes that preceding interactions change the current retrieval problem.  QReCC couples question rewriting, passage retrieval, and reading over multi-turn histories~\cite{anantha2021qrecc}.  ConvDR encodes conversational context directly~\cite{yu2021convdr}, and COTED removes distracting history while training a conversational dense retriever~\cite{mao2022coted}.  CONQRR instead learns a context-dependent rewrite that improves an existing retriever~\cite{wu2022conqrr}.  Long-term benchmarks such as LoCoMo and LongMemEval test cross-session reasoning, temporal memory, updates, and abstention~\cite{maharana2024locomo,wu2025longmemeval}.  These protocols define the history supplied to a current turn, but they do not pair one anchor under PIT and Future interaction states or require a certificate that timestamps each selected memory entry.  This distinction preserves their current-turn conclusions while leaving a historical system-state claim unidentified.  PIT Discovery supplies the paired replay and entry-level audit.

\textbf{Temporal retrieval and evaluation.}
Temporal IR models time in queries, documents, events, and relevance~\cite{kanhabua2015temporal}.  Recent benchmarks add several temporal settings.  StreamingQA evaluates quarterly adaptation to an expanding news corpus~\cite{liska2022streamingqa}; RealTime QA issues new questions every week and retrieves current evidence~\cite{kasai2023realtimeqa}; and Wild-Time evaluates learning under timestamped distribution shift~\cite{yao2022wildtime}.  FreshStack constructs realistic technical-document retrieval tasks with query dates~\cite{thakur2025freshstack}, which we use for natural-order validation.  Luu et al. measure training--test temporal misalignment~\cite{luu2022timewaits}, while LongEval varies training and test collections to measure retrieval persistence~\cite{cancellieri2025longeval}.  These studies quantify corpus, model, query, or distribution change.  Our counterfactual freezes those variables and changes only the eligibility time of interaction memory.

\begin{table}[t]
\centering
\caption{Adjacent evaluation lines vary different state components.  PIT Discovery isolates memory time under a fixed corpus and index.}
\label{tab:related_axes}
\footnotesize
\setlength{\tabcolsep}{2.3pt}
\begin{tabular}{@{}p{1.42cm}p{1.35cm}p{1.55cm}p{2.58cm}@{}}
\toprule
Line & Corpus time & Interaction state & Primary question \\
\midrule
Data discovery & Fixed & Absent & Which assets form the product? \\
Conv. IR & Fixed & Given history & Does context improve the current turn? \\
Temporal QA/shift & Varies & Usually absent & Does quality persist as data evolve? \\
Rec. leakage & Often co-varies & Training/user history & Does the split leak future events? \\
PIT Discovery & Fixed & Paired by time & How much score comes from future memory? \\
\bottomrule
\end{tabular}
\end{table}

\textbf{Leakage in offline recommendation.}
Recommendation supplies the closest causal precedent.  Ji et al. impose a global timeline on training interactions and item availability and show changes in accuracy and model order~\cite{ji2023leakage}.  Hidasi and Czapp catalogue recurring offline-evaluation flaws~\cite{hidasi2023flaws}.  Two 2025 studies then compare temporal split policies and quantify leakage in sequential recommendation~\cite{gusak2025split,le2025ahead}.  Broader audits connect leakage to irreproducible scientific claims~\cite{kapoor2023leakage} and motivate stronger reporting standards~\cite{jannach2026standards}.  These recommender analyses jointly involve training data, user histories, item availability, or split policy.

PIT Discovery builds on their causal principle and isolates a narrower state variable.  The paired replay holds the asset corpus, index, anchor, base ranking, reranker, and budgets fixed.  It changes only the temporal eligibility of query-trace memory and reports Temporal Violation Rate beside retrieval quality.  The contribution is a controlled measurement and audit protocol for a historical memory snapshot, without a priority claim for temporal leakage itself.

\section{Point-in-Time Discovery}
\label{sec:problem}

\subsection{A corpus snapshot is only part of the state}

Let $D_t$ be the set of discoverable assets frozen at historical corpus time $t$, and let $q_i$ be the request arriving at position $i$ in an interaction stream.  A stateless retriever appears to be fully specified by $(q_i,D_t,\theta_t)$, where $\theta_t$ contains its fixed parameters and index.  A stateful system additionally consults a memory $M_i$ induced by prior retrieval events.  Its output is
\begin{equation*}
R_i=R(q_i;D_t,\theta_t,M_i,B),
\end{equation*}
where $B$ fixes retrieval and memory-access budgets.

A historically valid system state is therefore not $D_t$ alone but
\begin{equation}
\mathcal{S}_{t,i}=\left(D_t,\theta_t,M_{<i}\right).
\label{eq:state}
\end{equation}
Here, $M_{<i}$ may contain only events observed strictly before the anchor.  Freezing $D_t$ while constructing $M_i$ from the full recorded stream combines yesterday's assets with tomorrow's experience.  The resulting system never existed at the claimed point in time.

This distinction is orthogonal to temporal relevance.  Temporal IR studies time-sensitive needs and documents~\cite{kanhabua2015temporal}, and longitudinal test collections measure performance as data evolve~\cite{cancellieri2025longeval}.  We instead hold the corpus and query fixed to isolate the availability time of state accumulated by the system.

\subsection{Three clocks define historical validity}

A stateful replay combines objects governed by three clocks.  The \emph{content clock} $c(a)$ records when asset $a$ enters the discoverable corpus.  The \emph{observation clock} $o(e_j)$ records when the system observes interaction event $e_j$.  The \emph{materialization clock} $\mu(M)$ records when an evaluator builds a cache, profile, trace index, or retrieval graph.  An anchor at $(t,i)$ requires entry-level eligibility
\begin{equation*}
\mathbf{1}[c(a)\leq t]\;\mathbf{1}[o(e_j)<o(q_i)] = 1
\end{equation*}
for every asset and memory event that influences its output.  A corpus filter enforces the first term; a PIT memory view enforces the second.

The materialization timestamp cannot certify the state by itself.  An evaluator may rebuild a cache after the study period and still obtain a valid PIT memory view if every entry carries an observation time and the view filters that time at the anchor.  Conversely, a cache built before evaluation may already contain post-anchor traces.  PIT Discovery therefore versions the materialization recipe but audits eligibility at the event level.

This separation narrows the causal claim.  We audit the dependency path from selected memory entries to the ranked output.  Pretraining data, latent model knowledge, and unselected cache records remain outside that path.  TVR measures the future share among entries the reranker selected, so a zero value certifies the implemented memory boundary for that output.

\subsection{Memory views}

Let the event log be $L=(e_1,\ldots,e_n)$, with event $e_j$ observed at position $j$ and containing a query plus its behavior-only retrieval trace.  For anchor $i$, we define four principal memory views:
\begin{align*}
M_i^{S} &= \varnothing && \text{(Stateless)},\\
M_i^{P} &= \{e_j:j<i\} && \text{(PIT)},\\
M_i^{F(h)} &= M_i^{P}\cup\{e_j:i<j\le i+h\} && \text{(Future-$h$)},\\
M_i^{F} &= \{e_j:j\ne i\} && \text{(Future)}.
\end{align*}
Every view excludes the anchor event.  A fifth, deliberately Naive condition freezes $D_t$ but materializes memory from all of $L$ before evaluation.  Under identical selection logic, Naive should equal Future; this equality is an end-to-end leak control.

Table~\ref{tab:protocol} separates information availability from ordinary retrieval choices.  PIT and Future have equal memory lookup budgets and differ only in which event timestamps are eligible.  This paired design is the identifying assumption of our measurement: once the evaluator fixes $q_i$, $D_t$, $\theta_t$, $B$, and the ranking rule, Future eligibility explains the score difference.

\begin{table}[t]
\centering
\caption{Memory views at anchor $i$.  The anchor's own trace is always excluded.}
\label{tab:protocol}
\small
\setlength{\tabcolsep}{4pt}
\begin{tabular}{lccc}
\toprule
Condition & Past $j<i$ & Future $j>i$ & Valid PIT? \\
\midrule
Stateless & -- & -- & \checkmark \\
PIT & \checkmark & -- & \checkmark \\
Future-$h$ & \checkmark & $j\le i+h$ & $\times$ \\
Future & \checkmark & \checkmark & $\times$ \\
Naive & \checkmark & \checkmark & $\times$ \\
\bottomrule
\end{tabular}
\end{table}

\subsection{The paired replay identifies memory-time bias}

Let $Y_i(c)$ denote a retrieval metric for anchor $i$ under memory condition $c$.  The observed pair $Y_i(F)-Y_i(P)$ compares two potential outputs for the same request.  Pairing removes query difficulty, product size, and base-ranking quality from the contrast because both branches reuse the same anchor and base top-200 list.  Averaging these within-query differences defines the reported Hindsight Gain.

The comparison also requires a stable event log.  We construct every behavior-only trace once from the fixed base retriever before condition-specific replay.  PIT and Future read that shared log but never write condition-dependent events back into it.  Otherwise an early treatment choice could alter the history available to a later anchor and mix direct future access with policy feedback.

Three conditions support the attribution.  First, both branches use identical retrieval and memory budgets.  Second, stream position is the sole eligibility input; relevance labels never select memory.  Third, the reranker is deterministic given the selected entries.  The protocol estimates the direct effect of memory-time eligibility under these conditions.  Long-run effects from branch-dependent user behavior remain outside this estimand.

\subsection{Measurement and validity}

For a quality metric $m_k$ at cutoff $k$, we define the primary quantity:
\begin{equation}
\mathrm{HG}_k=m_k(R_i^{F})-m_k(R_i^{P}),
\label{eq:hg}
\end{equation}
the \emph{Hindsight Gain}.  HG credits the score difference to information unavailable at the historical anchor.  Future-$h$ produces a dose--response curve that shows how rapidly the bias enters.

We pair quality with an explicit validity metric.  If $A_i^c$ is the set of memory events selected under condition $c$, then
\begin{equation}
\mathrm{TVR}_i^c=\frac{1}{|A_i^c|}\sum_{e_j\in A_i^c}\mathbf{1}[j>i],
\label{eq:tvr}
\end{equation}
with TVR set to zero for empty memory.  PIT is valid only if TVR is identically zero, not merely small on average.

Two usual utility ratios help only when memory improves over Stateless: Hindsight Share divides HG by $|m(F)-m(S)|$, while PIT Utility Retention compares $(m(P)-m(S))$ with $(m(F)-m(S))$.  Our experiments show a more difficult regime in which both memories can hurt.  We therefore report the signed PIT memory effect and HG, then combine their paired query-cluster means in the masked-harm fraction of Section~\ref{sec:result_mask}.  This avoids describing recovered damage as ``retained utility.''

\paragraph{Scope of the estimand.}
The protocol estimates hindsight inflation for the specified memory mechanism and stream distribution.  Training-data contamination, corpus drift, user satisfaction, and effects on unobserved online policies remain outside the estimand.  The protocol tells whether an offline score answers the deployable PIT question.

\section{The PIT Discovery Protocol}
\label{sec:protocol}

PIT Discovery turns Eq.~\ref{eq:state} into a six-stage replay: immutable assets, deterministic streams, online traces, paired memory views, paired metrics, and temporal audit.  The protocol enforces four invariants: immutable inputs, label-free event construction, paired memory views, and machine-checkable temporal validity.

\subsection{Immutable assets and typed products}

The input is a hybrid corpus and a set of requests with relevant asset identifiers.  We normalize tables and passages into a single asset namespace while retaining their types.  Each request defines a \emph{product}: the set of table and text assets needed to satisfy it.  This representation supports both asset-level recall and stricter product-level metrics without inserting answer generation into the evaluation.

Every transformation emits a manifest.  The corpus content, serialized assets, type vector, queries, embedding files, and retrieval outputs each receive cryptographic hashes.  The checker rejects a paired run when its PIT and Future rows lack shared corpus and index hashes.  Hashes expose accidental snapshot mismatch and provide no adversarial security guarantee.

\subsection{Deterministic query streams}

Because interaction produces memory, an unordered query set is insufficient.  We convert the same requests into deterministic arrival streams.  A random permutation measures diffuse access.  A semantic-burst stream groups nearby query embeddings and emits short within-topic runs, modeling workloads in which related analytical requests arrive together.  The manifest records seeds and stream hashes, and each query occurs once per stream.

Headline stream construction forbids relevance labels.  This matters because grouping queries by shared gold assets would manufacture ideal conditions for trace reuse.  The pipeline tags any such stress stream with \texttt{uses\_gold=true}; aggregation excludes it from natural-stream claims.

\subsection{Online trace construction}

We first run the fixed base retriever for every query.  The system then traverses the stream once in arrival order.  At position $j$, it writes an event
\begin{equation*}
e_j=(q_j,\,\pi_j^{1:d},\,s_j^{1:d},\,j,\,v),
\end{equation*}
where $\pi_j^{1:d}$ and $s_j^{1:d}$ are the top-$d$ retrieved assets and scores, $j$ is \texttt{observed\_at}, and $v$ hashes the retrieval source.  In our implementation $d=50$.  This behavior-only memory captures retriever exposure without recording relevance.

The log is append-only.  Constructing it online prevents the implementation from accidentally attaching the anchor's gold product or a future-derived representation to a past event.  We separately store the query-to-position and position-to-query maps needed to materialize views efficiently.

For the revenue request in Section~\ref{sec:intro}, its product remains the same table--passage set in every condition.  The causal branch may reuse only earlier request traces; the following week's similar request enters only Future.  This is the end-to-end distinction illustrated in Figure~\ref{fig:main}.

\subsection{Paired replay at an anchor}

At anchor $i$, the evaluator loads the base top-200 list once, sorts query-neighbor candidates by embedding similarity, and applies the eligibility predicate of each condition in Table~\ref{tab:protocol}.  PIT and Future both select at most 32 entries, preserving equal capacity.  The same deterministic reranker combines trace evidence with the base list.

Each output row records the anchor, cutoff, selected memory positions, memory and retrieval budgets, stream seed, corpus hash, index hash, quality metrics, and TVR.  This row-level provenance makes the counterfactual inspectable: a reviewer can identify precisely which future events produced a nonzero TVR and recompute the paired score difference.

\subsection{Audit checks}

Three tests close common loopholes.  \textbf{Causal eligibility}: every selected PIT position must be less than the anchor.  \textbf{Self exclusion}: Future also disallows position $i$, preventing the anchor from retrieving its own stored result.  \textbf{Naive identity}: full-stream Naive must match explicit Future for selected entries and output metrics.  Together these tests distinguish a genuine memory-time effect from query duplication, budget drift, or inconsistent implementations.

Table~\ref{tab:audit_contract} turns these tests into an artifact contract.  Every paired row must carry enough provenance for an independent checker to reject a mismatched corpus, altered budget, post-anchor PIT entry, or inconsistent full-stream control.  The checker treats a violation as an invalid run instead of averaging it into a quality score.

\begin{table}[t]
\centering
\caption{Machine-checkable replay contract.  Any failed invariant invalidates the paired row.}
\label{tab:audit_contract}
\footnotesize
\setlength{\tabcolsep}{3.0pt}
\begin{tabular}{@{}p{1.35cm}p{3.05cm}p{2.75cm}@{}}
\toprule
Object & Required equality or order & Recorded certificate \\
\midrule
Corpus & Same assets and types & Corpus hash \\
Index & Same embeddings and base ranks & Index/source hash \\
Anchor & Same query, product, and cutoff & Query ID, $k$ \\
Budget & Same trace depth and slots & $d$, neighbor cap \\
PIT view & Every selected $j<i$ & Selected positions, TVR \\
Control & Naive output equals Future & Row-level max difference \\
\bottomrule
\end{tabular}
\end{table}

The contract also separates data errors from scientific outcomes.  A negative PIT memory effect remains a valid result because the row passes every causal check.  A positive score with TVR above zero fails the historical claim even when all retrieval metrics improve.  This distinction prevents quality from serving as evidence of temporal validity.

The protocol is intentionally independent of the memory internals.  A learned memory, cache, profile, or graph can replace our trace reranker when its entries have observable availability times and the evaluator can materialize paired views under the same capacity.  The contribution centers on a state and evaluation contract.

\section{Experimental Setup}
\label{sec:setup}

We design the experiments as paired counterfactuals.  Every pair fixes the corpus, index, anchor, base ranking, reranker, and budgets across PIT and Future.  This section specifies the shared infrastructure and statistical analysis.

\subsection{Benchmark and corpus construction}

We use the public DPDisc release of DPBench~\cite{zhang2025dpbench}, which converts table--text question-answering datasets into requests whose answer requires discovering a set of data assets.  ConvFinQA~\cite{chen2022convfinqa} and HybridQA~\cite{chen2020hybridqa} are the two core domains; we reserve TAT-QA~\cite{zhu2021tatqa} as a robustness domain.  We concatenate the public train, development, and test requests because labels support evaluation only; they never order streams, build memory, or rerank assets.

Each corpus object becomes an immutable typed asset: a minimally serialized table or a title--passage pair.  Query labels are lists of relevant table and text asset identifiers.  Table~\ref{tab:data} summarizes the resulting snapshots.  Every relevant identifier maps to one asset; there are no missing or empty gold products.  Snapshot manifests record source, asset, type-array, and query-file SHA-256 hashes.

\begin{table}[t]
\centering
\caption{DPDisc snapshots used in the replay.  Gold is the mean number of relevant assets per request.}
\label{tab:data}
\small
\setlength{\tabcolsep}{3.6pt}
\begin{tabular}{lrrrrr}
\toprule
Domain & Queries & Tables & Text & Assets & Gold \\
\midrule
ConvFinQA & 3,113 & 4,976 & 8,721 & 13,697 & 13.16 \\
HybridQA  & 8,820 & 12,378 & 41,608 & 53,986 & 30.61 \\
TATQA     & 1,143 & 2,757 & 4,760 & 7,517 & 10.52 \\
\bottomrule
\end{tabular}
\end{table}

\subsection{Retrievers and trace memory}

\textbf{Base retrieval.}  We test BM25 ($k_1=1.5$, $b=0.75$)~\cite{robertson2009bm25} and a hybrid retriever.  The dense component embeds assets and requests with \texttt{all-mpnet-base-v2}, using normalized cosine similarity in the Sentence-BERT framework~\cite{reimers2019sbert}.  Hybrid retrieval applies Reciprocal Rank Fusion to the top-200 BM25 and dense lists with constant 60.  All conditions receive the same top-200 base list.  Before replay, hybrid Recall@100 is 78.83\% for ConvFinQA, 38.89\% for HybridQA, and 63.94\% for TATQA.  At least one gold asset appears in the top 100 for 97.65\%, 89.86\%, and 96.68\% of requests, respectively.

\textbf{Memory vehicle.}  A transparent, behavior-only trace memory makes the measured mechanism inspectable.  After each stream query, it stores the top 50 assets and raw scores returned by the fixed base retriever, together with query identity, observation position, and source version.  It never stores relevance labels or answers.  For an anchor $q_i$, we scan its 1,024 nearest query neighbors, retain the first 32 allowed by the condition, and aggregate their trace evidence.  If neighbor $j$ has cosine similarity $s_{ij}$ and asset $a$ occurs at trace rank $r_{ja}$, its unnormalized memory score is
\begin{equation*}
u_i(a)=\sum_{j\in N_i}\frac{s_{ij}^{2}}{\log_2(r_{ja}+2)}.
\end{equation*}
After max normalization, we add $0.2u_i(a)$ to the reciprocal base rank $1/(r_i(a)+1)$ and return the top 200 assets.  This mechanism is an evaluation instrument: its simplicity makes the provenance of every score change inspectable.  We make no competitive retrieval claim for it.

\textbf{Useful-memory robustness.}  To test whether the effect depends on a harmful trace rule, we simulate a positive-feedback cache.  After a query, its relevant assets stand in for perfectly observed clicks; related queries contribute the maximum squared query similarity for each clicked asset.  We cap the strongest bonus at $0.02$, the reciprocal score at base rank 50, without tuning it on outcomes.  This label-based oracle proxy serves only as a mechanism stress test; it asks whether a memory that helps under PIT remains vulnerable to future access.

\subsection{Streams, anchors, and conditions}

DPDisc does not publish real discovery logs, so we construct two label-free arrival regimes.  We seed each \emph{Random} permutation.  \emph{Semantic} streams cluster request embeddings with MiniBatch $k$-means and emit short topical bursts of 4--16 requests.  These streams approximate locality without consulting gold assets.  We generate five seeds of each regime per domain.  A gold-overlap ordering exists only as a code-level stress test, and we exclude it from every reported result.

For each stream we select 100 anchors around each of the 20\%, 40\%, 60\%, and 80\% positions, producing 400 anchor evaluations.  The complete design has three domains, two streams, two retrievers, five seeds, 400 anchors, and nine memory conditions, for 216,000 query-condition rows.  Conditions are Stateless, PIT, Future, Naive, and Future-$h$ for $h\in\{10,50,100,250,500\}$.  Every condition excludes the anchor's own trace, preventing a trivial exact-query leak.  Each headline cell contains 2,000 paired anchor observations.

We also run a natural-order study on FreshStack, a timestamped technical-document retrieval benchmark~\cite{thakur2025freshstack}.  Its Angular, Godot, LangChain, Laravel, and YOLO topics contain 672 queries and 271,816 text assets, with no missing relevance identifiers.  We sort each topic by its recorded query time and evaluate every position with at least eight past and eight future events, giving 592 equal-capacity anchors.  This study uses BM25 retrieval and query-local TF--IDF similarity only for trace selection; neither stream order nor memory construction uses relevance labels.

\subsection{Metrics and uncertainty}

The headline metric is Asset Recall@100 over the union of relevant table and text assets.  We also report Recall@$k$ for $k\in\{10,20,50\}$, type-specific recall, product coverage, complete-product success, and the rank needed to inspect 80\% of a product.  Our paired diagnostics are Hindsight Gain (Eq.~\ref{eq:hg}), the PIT memory effect $\mathrm{Score}(P)-\mathrm{Score}(S)$, and the masked-harm fraction in Eq.~\ref{eq:mhf}.  Temporal Violation Rate audits availability; retrieval metrics quantify quality.

Repeated seeds can expose the same request at different anchors.  Within each matched anchor, we form $HG_i=Y_i(F)-Y_i(P)$ and $\Delta_{\mathrm{PIT},i}=Y_i(P)-Y_i(S)$.  We average each contrast by query identifier, then bootstrap query clusters with 5,000 resamples.  MHF uses the resulting query-cluster means, so every headline value follows one paired estimator.  The intervals quantify uncertainty across sampled query clusters under each tested workload.  Claims about unseen organizations or deployment streams require a new paired replay.

\subsection{Implementation and reproducibility}

We implement the pipeline in Python with NumPy, SciPy, scikit-learn, PyTorch, and Sentence Transformers.  Embedding and neighbor search use four RTX~3090 GPUs; replay and bootstrap analysis run on CPU.  Deterministic manifests record stream seeds, retrieval and memory hashes, budgets, selected memory positions, and output hashes for each run.  Fifteen unit tests cover schema validity, natural ordering, budget parity, self-trace exclusion, gold-label isolation, and temporal invariants.  Unit tests require equality between full-stream Naive and explicit Future; PIT runs abort on nonzero TVR.

\section{Results}
\label{sec:results}

We organize the results around the question a historical replay should answer: what could the system have retrieved \emph{then}?  Unless noted otherwise, every point estimate uses the paired query-cluster estimator in Section~\ref{sec:setup}.  Table~\ref{tab:main} reports the primary paired comparison.

\subsection{Future systematically inflates historical recall}
\label{sec:result_hg}

\textbf{Future exceeds PIT in every cell.}
Across all 12 combinations of domain, stream, and retriever, Future scores higher than PIT.  The gain in Asset Recall@100 ranges from \textbf{2.62 to 5.24 percentage points}; every paired interval excludes zero.  Matching the anchor query, frozen asset snapshot, base ranking, reranking rule, and budgets rules out corpus and index effects.  The only difference is whether the 32-query memory lookup may select observations after the anchor.

The effect spans both ends of the quality range.  ConvFinQA with semantic streams and BM25 has the largest HG at 5.24 points.  The harder HybridQA corpus with random streams and BM25 has the smallest HG at 2.62 points.  Hindsight therefore persists across corpora with different retrieval difficulty.

\begin{table*}[t]
\centering
\caption{Primary paired Asset Recall@100 effects.  HG forms Future$-$PIT within each anchor, then averages by query identifier.  Values are points with query-cluster bootstrap 95\% intervals; all exclude zero.}
\label{tab:main}
\setlength{\tabcolsep}{4.0pt}
\small
\begin{tabular}{llll}
\toprule
Domain & Stream & Retriever & HG [95\% CI] \\
\midrule
\multirow{4}{*}{ConvFinQA}
 & Random   & BM25   & $+4.79$ [4.18, 5.40] \\
 & Random   & Hybrid & $+4.92$ [4.31, 5.54] \\
 & Semantic & BM25   & $+5.24$ [4.54, 5.94] \\
 & Semantic & Hybrid & $+4.92$ [4.31, 5.53] \\
\midrule
\multirow{4}{*}{HybridQA}
 & Random   & BM25   & $+2.62$ [2.32, 2.94] \\
 & Random   & Hybrid & $+2.75$ [2.41, 3.07] \\
 & Semantic & BM25   & $+2.85$ [2.50, 3.19] \\
 & Semantic & Hybrid & $+2.80$ [2.46, 3.15] \\
\midrule
\multirow{4}{*}{TATQA}
 & Random   & BM25   & $+5.13$ [4.55, 5.74] \\
 & Random   & Hybrid & $+4.69$ [4.11, 5.29] \\
 & Semantic & BM25   & $+4.97$ [4.34, 5.62] \\
 & Semantic & Hybrid & $+4.69$ [4.07, 5.33] \\
\bottomrule
\end{tabular}
\end{table*}

\subsection{Leakage masks a harmful memory mechanism}
\label{sec:result_mask}

The paired utility analysis exposes a more consequential failure than an optimistic score.  \textbf{The paired PIT memory effect is negative in all 12 cells}: $-14.37$ to $-6.57$ recall points.  Trace reuse therefore contains no hidden improvement; for this deliberately simple, behavior-only memory, it displaces strong base-retrieval candidates with assets inherited from related but non-identical queries.  This negative result is important because it is what a deployable replay would show.

Future traces partially conceal that failure.  Define the \emph{masked-harm fraction} as
\begin{equation}
\mathrm{MHF}=\frac{\mathrm{Score}(F)-\mathrm{Score}(P)}{|\mathrm{Score}(P)-\mathrm{Score}(S)|},
\label{eq:mhf}
\end{equation}
when PIT is below Stateless.  Future information masks \textbf{32.7--48.4\%} of the deployable memory harm.  The ConvFinQA semantic stream with BM25 reaches 48.4\% from the unrounded paired estimates.  Even Future remains below Stateless in all cells, but it makes an unreliable mechanism look materially less harmful.  A model-selection process that sees only full-stream memory can consequently retain, tune, or deploy the wrong component.

This finding also sharpens the interpretation of HG.  Our claim concerns evaluation validity across harmful and useful memory mechanisms.  For the transparent trace-memory vehicle, hindsight changes the empirical judgment about the same mechanism.

\subsection{Useful memory remains vulnerable to hindsight}
\label{sec:result_feedback}

The positive-feedback proxy closes the missing utility quadrant.  In all 12 domain/stream/retriever cells, PIT improves on Stateless and Future improves again on PIT: $S<P<F$.  PIT utility ranges from \textbf{4.65 to 18.96 points}, while HG ranges from \textbf{4.11 to 9.58 points}; every query-cluster interval for both differences excludes zero.  Table~\ref{tab:feedback} groups the four settings per domain.  The perfect-click assumption makes this proxy a mechanism stress test.  Future leakage survives when prior feedback improves PIT retrieval, which rules out a deliberately weak trace reranker as the sole explanation.

\begin{table}[t]
\centering
\caption{Positive-feedback robustness (Recall@100 points).  Ranges span two streams and two retrievers; all 24 paired intervals exclude zero.}
\label{tab:feedback}
\small
\setlength{\tabcolsep}{5pt}
\begin{tabular}{lcc}
\toprule
Domain & PIT utility $P-S$ & HG $F-P$ \\
\midrule
ConvFinQA & 4.65--8.26 & 4.11--5.91 \\
HybridQA  & 13.08--18.96 & 5.03--7.39 \\
TATQA     & 13.10--16.15 & 8.30--9.58 \\
\bottomrule
\end{tabular}
\end{table}

\subsection{Leakage begins locally in semantic streams}
\label{sec:result_horizon}

\begin{figure*}[t]
  \centering
  \begin{tikzpicture}
\begin{groupplot}[
  group style={group size=3 by 2, horizontal sep=0.72cm, vertical sep=0.95cm},
  width=0.325\textwidth,
  height=3.45cm,
  tick label style={font=\fontsize{8}{9}\selectfont},
  label style={font=\fontsize{8.5}{9.5}\selectfont},
  title style={font=\fontsize{9.5}{10.5}\selectfont\bfseries},
  grid=major,
  grid style={draw=gray!18},
  axis line style={draw=gray!65},
  tick align=outside,
  legend style={font=\fontsize{8.5}{9.5}\selectfont, draw=none, fill=none,
    /tikz/every even column/.append style={column sep=5pt}},
  legend columns=2,
]
\nextgroupplot[
  title={ConvFinQA}, ylabel={HG (points)}, xlabel={Future horizon},
  symbolic x coords={10,50,100,250,500,Full}, xtick=data,
  ymin=-0.15, ymax=6.0, ytick={0,1,2,3,4,5,6}, legend to name=horizonlegend]
  \addplot+[mark=*, thick, color=pitblue] coordinates {(10,0.8023) (50,1.3360) (100,1.4670) (250,2.0561) (500,2.7341) (Full,4.9170)};
  \addlegendentry{Semantic}
  \addplot+[mark=square*, dashed, thick, color=leakred] coordinates {(10,0.0395) (50,0.1733) (100,0.3692) (250,1.0416) (500,2.1959) (Full,4.9215)};
  \addlegendentry{Random}

\nextgroupplot[
  title={HybridQA}, ylabel={}, xlabel={Future horizon},
  symbolic x coords={10,50,100,250,500,Full}, xtick=data,
  ymin=-0.15, ymax=6.0, ytick={0,1,2,3,4,5,6}]
  \addplot+[mark=*, thick, color=pitblue] coordinates {(10,0.6500) (50,0.9328) (100,1.0037) (250,1.0755) (500,1.1798) (Full,2.7990)};
  \addplot+[mark=square*, dashed, thick, color=leakred] coordinates {(10,0.0258) (50,0.0509) (100,0.1130) (250,0.2019) (500,0.5683) (Full,2.7459)};

\nextgroupplot[
  title={TATQA}, ylabel={}, xlabel={Future horizon},
  symbolic x coords={10,50,100,250,500,Full}, xtick=data,
  ymin=-0.15, ymax=6.0, ytick={0,1,2,3,4,5,6}]
  \addplot+[mark=*, thick, color=pitblue] coordinates {(10,1.3640) (50,2.2755) (100,2.6623) (250,3.7911) (500,4.4908) (Full,4.6907)};
  \addplot+[mark=square*, dashed, thick, color=leakred] coordinates {(10,0.0393) (50,0.5088) (100,1.1548) (250,2.3500) (500,3.8745) (Full,4.6949)};

\nextgroupplot[
  ylabel={HG (points)}, xlabel={Anchor percentile},
  symbolic x coords={20,40,60,80}, xtick=data,
  ymin=-0.2, ymax=11.0, ytick={0,2,4,6,8,10}]
  \addplot+[mark=*, thick, color=pitblue] coordinates {(20,7.1346) (40,5.7708) (60,4.7931) (80,1.9295)};
  \addplot+[mark=square*, dashed, thick, color=leakred] coordinates {(20,10.0611) (40,5.6610) (60,2.4345) (80,1.0299)};

\nextgroupplot[
  ylabel={}, xlabel={Anchor percentile},
  symbolic x coords={20,40,60,80}, xtick=data,
  ymin=-0.2, ymax=11.0, ytick={0,2,4,6,8,10}]
  \addplot+[mark=*, thick, color=pitblue] coordinates {(20,5.2761) (40,3.0052) (60,1.6692) (80,1.0767)};
  \addplot+[mark=square*, dashed, thick, color=leakred] coordinates {(20,5.2162) (40,2.8765) (60,1.6463) (80,0.6649)};

\nextgroupplot[
  ylabel={}, xlabel={Anchor percentile},
  symbolic x coords={20,40,60,80}, xtick=data,
  ymin=-0.2, ymax=11.0, ytick={0,2,4,6,8,10}]
  \addplot+[mark=*, thick, color=pitblue] coordinates {(20,6.6416) (40,6.4012) (60,3.9997) (80,2.1626)};
  \addplot+[mark=square*, dashed, thick, color=leakred] coordinates {(20,8.1029) (40,5.6566) (60,3.6783) (80,1.5011)};
\end{groupplot}
\node at ($(group c2r2.south)+(0,-1.22cm)$) {\pgfplotslegendfromname{horizonlegend}};
\end{tikzpicture}
  \caption{Hindsight Gain for the hybrid retriever.  Top: HG as the future horizon expands.  Bottom: HG by anchor position.  Solid lines are semantic streams; dashed lines are random streams.  The position trend remains despite 100\% slot fill in both memory views.}
  \Description{Six line charts show Hindsight Gain by future horizon and anchor percentile for ConvFinQA, HybridQA, and TATQA. Semantic streams rise immediately with horizon, while hindsight gain falls toward later anchors.}
  \label{fig:horizon}
\end{figure*}
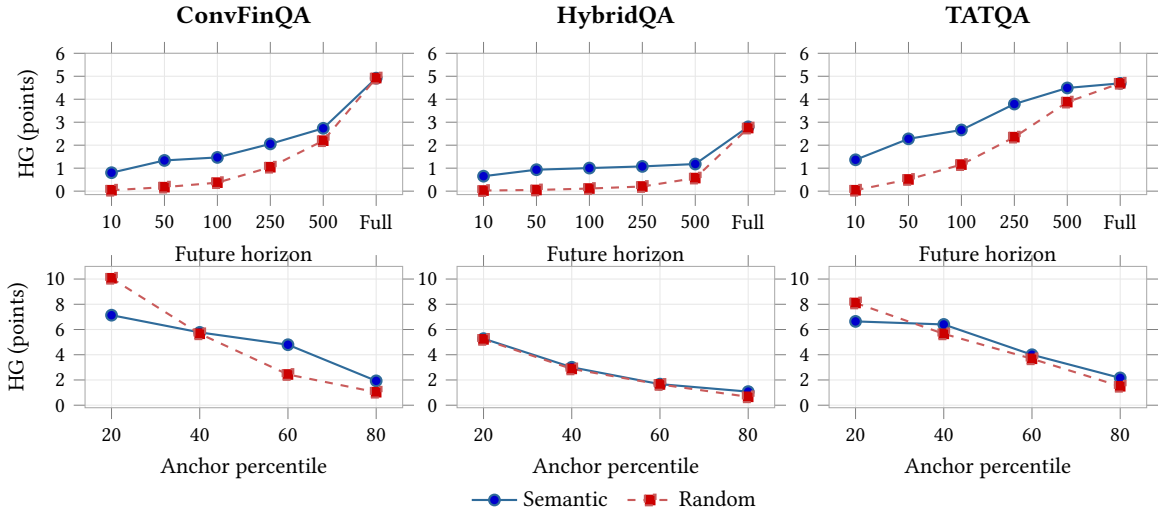

Figure~\ref{fig:horizon} doses the leak by future horizon.  In semantic-burst streams, only ten future queries already add 0.65--1.36 recall points with the hybrid retriever, and all three domain-level intervals exclude zero.  In random streams, none of the three horizon-10 intervals excludes zero.  The contrast localizes the mechanism: a nearby future query is useful when arrival order preserves topical neighborhoods, while a random next query rarely supplies a related trace.

The top row of Figure~\ref{fig:horizon} saturates at different windows.  At horizon 500, semantic streams recover 56\% of the full HG on ConvFinQA, 42\% on HybridQA, and 96\% on TATQA.  TATQA therefore concentrates most leakage locally, whereas ConvFinQA and especially HybridQA retain a long tail.  The evidence rejects a universal ``small look-ahead is safe'' rule.

\textbf{Anchor position.}  The bottom row addresses an alternative explanation: early anchors have more future candidates.  Averaged over all 12 primary cells, HG declines from 7.26 points at the 20\% anchor to 1.43 at 80\%.  Yet PIT and Future fill all 32 slots at every quartile.  The selected future-slot share falls from 77.5\% to 25.7\%, while the nearest Future similarity remains 0.904--0.908 and PIT similarity rises from 0.853 to 0.900.  Complete slot fill and stable neighbor quality rule out capacity collapse; every quartile retains positive HG.

\subsection{The effect survives design changes}
\label{sec:result_robust}

\textbf{Retriever.}  BM25 and hybrid retrieval show nearly the same qualitative effect.  Within each domain and stream, switching the base retriever changes HG by at most 0.44 points.  Similar HG under BM25 rules out dense retrieval as the sole cause.  Hybrid improves the Stateless baseline on HybridQA while leaving Future causally invalid.

\textbf{Domain.}  TATQA is a held-out robustness domain in our design, and its 4.69--5.13 point gains fall inside the range observed on the two core domains.  The result spans Wikipedia-style hybrid evidence and financial table--text evidence.

\textbf{Natural arrival order.}  Across 592 balanced FreshStack anchors, Future exceeds PIT by \textbf{2.72 points} [1.75, 3.71].  PIT TVR remains zero, and Future equals Naive.  Here legal trace memory is 4.58 points below Stateless, so the result also reinforces the distinction between utility and validity.  The positive HG under recorded query timestamps extends the effect beyond our random and semantic-burst generators.

\textbf{Metric.}  Future improves Asset Recall at every cutoff: HG ranges from 0.32--2.02 points at $k=10$, 0.66--3.71 at $k=20$, and 1.68--4.73 at $k=50$.  At $k=100$, table recall rises by 1.45--8.70 points and text recall by 2.85--5.93 points.  Future also reduces the number of files inspected to reach 80\% product coverage by 0.39--5.69.  Complete-product success is directionally positive, although the smallest effects are too sparse for a uniformly nonzero interval.  The headline conclusion therefore reflects better retrieval of both asset types.

\textbf{Near duplicates.}  We removed every anchor for which either PIT or Future selected a query neighbor with cosine similarity at least 0.99.  This retained 99.9--100.0\% of observations; HG remained 2.62--5.24 points, and all 12 intervals still excluded zero.  The effect therefore persists after excluding exact and near-exact anchor variants.

\subsection{The audit detects the exact invalid state}
\label{sec:result_audit}

The validity checks identify the invalid state.  PIT has a maximum Temporal Violation Rate of zero over all 24,000 PIT anchor evaluations.  Future selects post-anchor entries in 50.1--54.0\% of its 32 memory slots on average.  The deliberately Naive condition freezes the corpus but pre-materializes memory from the full stream.  It matches explicit Future on every metric and every query (maximum absolute difference $=0$).  This identity is a strong end-to-end control: future trace access reproduces the inflation from the common naive replay, while the zero-TVR invariant rejects it immediately.

Taken together, the results close the causal loop.  A fixed historical corpus can yield different offline conclusions because the replay leaves system state unfixed.  Pairing $D_t$ with $M_{<i}$ removes that ambiguity; reporting quality together with TVR exposes any hindsight credited as progress.

\section{Discussion}
\label{sec:discussion}

\textbf{Historical validity follows dependency provenance.}
A retrieval score describes a historical system only when its corpus, model, and interaction state were available at the anchor.  Benchmarks should publish memory-materialization rules and entry-level observation times, then report TVR beside quality.  We audit selected trace entries directly.  For any derived state object $m$, a conservative extension is:
\begin{equation*}
\max_{e\in\operatorname{Ancestors}(m)} o(e) < o(q_i).
\end{equation*}
This timestamps $m$ by its latest ancestor.  Systems can implement the condition with provenance pointers or a conservative upper-bound timestamp.  Coarse or tied timestamps require a conservative rule: the evaluator admits an event only when transaction metadata proves that it precedes the anchor.  Derived state inherits the latest possible observation time among its ancestors, which prevents an arbitrary within-batch order from creating false eligibility.

\textbf{Model and memory updates require separate versions.}
Our experiments keep $\theta_t$ fixed and vary $M_i$; production systems may update both.  Shared checkpoints, adapters, indexes, or profiles belong in $\theta_t$; request-time caches or retrieved history belong in $M_i$.  Both require versions, availability intervals, and dependency provenance showing that every contributing event precedes the anchor.  A checkpoint timestamp records materialization time but omits the event lineage that shaped its parameters.  Learned memories therefore need matched PIT and Future versions plus links from each evaluated state to timestamped training events.  A black-box service without this lineage can compare outcomes, but it cannot certify its hidden state as historical.

\textbf{Validity and utility answer different decisions.}
Sections~\ref{sec:result_mask} and~\ref{sec:result_feedback} show that future access can mask a harmful component or overstate a useful one.  The same leak can therefore preserve a component that PIT evidence supports removing or exaggerate one that PIT evidence supports retaining.  The audit contract gates historical validity, the interval for $\Delta_{\mathrm{PIT}}=P-S$ determines PIT utility, and $HG=F-P$ quantifies the direction and magnitude of contamination (Table~\ref{tab:decision_matrix}).

\begin{table}[t]
\centering
\caption{The audit gates validity before interpreting PIT utility.}
\label{tab:decision_matrix}
\footnotesize
\setlength{\tabcolsep}{3.4pt}
\begin{tabular}{@{}p{1.55cm}p{1.35cm}p{4.35cm}@{}}
\toprule
Signal & Criterion & Interpretation and action \\
\midrule
Audit contract & any failure & Historical claim fails; rebuild from the event log. \\
Audit contract & all pass & Admit the PIT score to utility analysis. \\
95\% CI of $\Delta_{\mathrm{PIT}}$ & lower $>0$ & Memory helps; retain it for the tested workload. \\
95\% CI of $\Delta_{\mathrm{PIT}}$ & upper $<0$ & Memory harms; reject or retune it. \\
95\% CI of $\Delta_{\mathrm{PIT}}$ & contains 0 & Utility is inconclusive; collect more evidence. \\
\bottomrule
\end{tabular}
\end{table}

\textbf{TVR detects temporal contamination independently of HG.}
HG measures the score effect of Future eligibility; TVR tests whether selected memory entries crossed the time boundary.  Future information can leave Recall@100 unchanged by moving irrelevant or below-cutoff assets or by offsetting another error.  A run with nonzero TVR remains historically invalid even when HG equals zero, while negative HG means contamination reduced the measured output.  The audit contract therefore gates admission; HG estimates the measured consequence.

\textbf{A compact release gate makes the audit operational.}
Before model selection, the evaluator freezes an anchor, corpus and model hashes, stream order, memory recipe, and budgets.  A valid run must preserve equal capacity, exclude the anchor's own trace, obtain row-level PIT TVR $=0$, and reproduce the Future$=$Naive leak control; query-paired intervals then accompany both signed components.  If a check fails, the evaluator rebuilds PIT and Future from one event log instead of patching an opaque cache.  This frozen-log design estimates the direct effect of memory eligibility.  Measuring long-run feedback additionally requires branch-specific logs because rankings can change later interactions.  The same audit contract applies to conversational search, recommendation, retrieval-augmented assistants, and tool-use agents when the system records availability and dependency provenance for its selected state.

\textbf{Repair starts from the event log.}
Dropping rows with future entries after scoring can change the query population and contaminate derived state.  A correct repair reconstructs every memory view from the shared append-only log, reapplies the eligibility predicate at each anchor, and reruns selection and ranking.  New hashes distinguish repaired outputs from the rejected materialization.  This procedure also preserves negative results: a legal PIT view may underperform, but it still answers the historical question.  The audit screens information availability while preserving unfavorable PIT results.

\textbf{The audit artifact makes the state boundary checkable.}
Each reported PIT result should bind the anchor identifier to corpus, model, and memory hashes in one manifest.  The same record should include the eligibility predicate, selected entry identifiers, observation times, dependency timestamps, and capacity budget.  Stable identifiers let evaluators protect query content while still recomputing TVR and checking capacity parity.  Reviewers can verify the boundary without rerunning embedding or retrieval because the manifest identifies the selected dependencies and their clocks.  Benchmarks can publish the schema and validation script beside scores so independent implementations test the same historical claim.  This artifact makes information availability a repeatable check.

\section{Limitations}
\label{sec:limitations}

Our effect sizes describe three DPDisc table--text domains, two retrievers, two transparent memory mechanisms, and five FreshStack topics.  Asset Recall measures evidence discovery; answer quality, user satisfaction, and downstream task success remain outside the evaluation.  The fixed-log replay isolates direct memory-time eligibility and excludes long-run branch-dependent behavior.  Query-cluster intervals cover repeated requests in the tested workloads; broader deployment claims require a new paired replay with auditable event lineage.  Systems that cannot expose timestamps or dependencies can undergo only a partial audit.

\section{Conclusion}
\label{sec:conclusion}

A historical corpus captures one part of a historical discovery system.  PIT Discovery pairs replay branches that hold the anchor, corpus, retriever, and budgets fixed while changing memory-time eligibility.  Across three DPDisc domains, Future raises Asset Recall@100 by 2.62--5.24 points in every primary cell; FreshStack shows the same direction under recorded order.  The two memory vehicles expose both model-selection errors: hindsight masks 32.7--48.4\% of trace-memory harm and overstates the benefit of positive-feedback memory.  TVR tests whether selected state is historical, while the signed PIT memory effect measures whether deployable memory helps.  Historical evaluation must bind $D_t$ to $M_{<i}$, version $\theta_t$ and derived state, and preserve observation and dependency timestamps in an auditable manifest.  This state contract makes historical scores independently verifiable.

\clearpage
\section*{Ethical Considerations}

This work evaluates retrieval over public research datasets and does not collect new personal data or involve human subjects.  The memory traces contain benchmark query identifiers and retrieved asset identifiers instead of user profiles.  The failure studied here has an ethical dimension: evaluations with post-anchor memory can overstate historical capability and encourage deployment of unreliable components.  We therefore report the negative PIT result, scope conditions, uncertainty, and validity checks without presenting leaked utility as progress.

The experiments used four RTX~3090 GPUs for embedding and nearest-neighbor computation; replay and analysis used CPUs.  We cached embeddings and base rankings across paired conditions to avoid repeated model inference.  Released manifests contain hashes and configuration metadata but exclude credentials and private production logs.  Practitioners applying the protocol to real logs should minimize retained query content, enforce access controls, and treat event-time provenance as sensitive operational metadata.  We used a large language model only to polish the manuscript's writing.

\bibliographystyle{ACM-Reference-Format}
\bibliography{references}

\end{document}